\documentclass[aps,pra,twocolumn,showpacs,groupedaddress]{revtex4}  
\usepackage{graphicx}  
\usepackage{dcolumn}   
\usepackage{bm}        
\usepackage{amssymb}   
\usepackage{draftcopy}

\begin{document}

\newcommand{\EuYSO}{Eu$^{3+}$:Y$_2$SiO$_5$~}
\newcommand{\Eu}{Eu$^{3+}$}
\newcommand{\YSO}{Y$_2$SiO$_5$}

\title{A cavity-stabilized laser with acceleration sensitivity below $10^{-12}$/g}

\author{David~R.~Leibrandt}
\email{david.leibrandt@nist.gov}
\author{James C. Bergquist}
\author{Till Rosenband}
\affiliation{National Institute of Standards and Technology, 325 Broadway St., Boulder, CO 80305, USA}

\date{\today}

\begin{abstract}
We characterize the frequency-sensitivity of a cavity-stabilized laser to inertial forces and temperature fluctuations, and perform real-time feed-forward to correct for these sources of noise.  We measure the sensitivity of the cavity to linear accelerations, rotational accelerations, and rotational velocities by rotating it about three axes with accelerometers and gyroscopes positioned around the cavity.  The worst-direction linear acceleration sensitivity of the cavity is $2(1) \times 10^{-11}$/g measured over 0--50~Hz, which is reduced by a factor of 50 to below $10^{-12}$/g for low-frequency accelerations by real-time feed-forward corrections of all of the aforementioned inertial forces.  A similar idea is demonstrated in which laser frequency drift due to temperature fluctuations is reduced by a factor of 70 via real-time feed-forward from a temperature sensor located on the outer wall of the cavity vacuum chamber.
\end{abstract}

\maketitle

\section{Introduction}

Frequency-stable lasers are indispensable tools for precision measurements, finding applications ranging from optical frequency standards \cite{Chou2010a,Huntemann2012,Madej2012} and tests of fundamental physics \cite{Rosenband2008,Eisele2009,Reynaud2009} to geodesy \cite{Chou2010b,Bondarescu2012} and low-phase-noise microwave synthesis \cite{Baumann2009,Zhang2010,Fortier2011,Benedick2012}.  Typically, lasers for such applications are stabilized by locking them to a Fabry-P\'{e}rot cavity such that the fractional frequency stability of the laser is determined by the fractional length stability of the cavity \cite{Drever1983,Young1999,Jiang2011,Kessler2012,Chen2012,Nicholson2012}.  Such cavity-stabilized lasers are sensitive to environmental perturbations that change the length of the cavity, including vibrations and temperature fluctuations, and much effort has gone into designing cavities that are less sensitive to accelerations \cite{Notcutt2005,Millo2009,Leibrandt2011a,Webster2011} and shielding them from temperature fluctuations \cite{Dawkins2009}.  For applications outside the laboratory, such as clock-based geodesy \cite{Chou2010b,Bondarescu2012}, it is desirable to further reduce the sensitivity of laser frequencies to environmental perturbations.

A complementary approach to the passive reduction of the vibration sensitivity of cavity-stabilized lasers has been demonstrated \cite{Thorpe2010,Leibrandt2011b} in which the acceleration environment is measured and real-time feed-forward is applied to correct the frequency of the laser for the acceleration-induced frequency fluctuations.  This approach also yields information about the transfer function that describes how accelerations change the length of the cavity, which can in turn be used to improve the cavity design to achieve a lower passive vibration sensitivity.  In previous work, feed-forward corrected for first-order deformations of the cavity due to linear and rotational accelerations over a bandwidth of 0.06--450~Hz.  Here, we extend this approach to include feed-forward that corrects for first- and second-order deformations of the cavity due to all of the (rigid body) inertial forces, namely rotational velocity as well as linear and rotational acceleration, with a bandwidth that extends to zero frequency (0--300~Hz).  In addition, we implement feed-forward to correct for temperature fluctuations of the cavity by use of a temperature sensor located on the cavity vacuum chamber.  This extended feed-forward is combined with a redesigned cavity mount that improves the passive acceleration sensitivity of the cavity to be equal to the best reported value for a Fabry-P\'{e}rot cavity \cite{Webster2011}.

This paper proceeds as follows.  Sec.~\ref{sec:setup} describes the experimental setup, including details of the cavity mount design and feed-forward implementation.  Measurements of the acceleration sensitivity of the cavity and the frequency stability of the laser in a laboratory environment are presented in Sec.~\ref{sec:stability}.  Finally, Secs.~\ref{sec:acceleration} and \ref{sec:temperature} evaluate the performance of inertial force and temperature feed-forward, respectively.

\section{Setup}\label{sec:setup}

\begin{figure}
\begin{center}
\includegraphics[width=0.9\columnwidth]{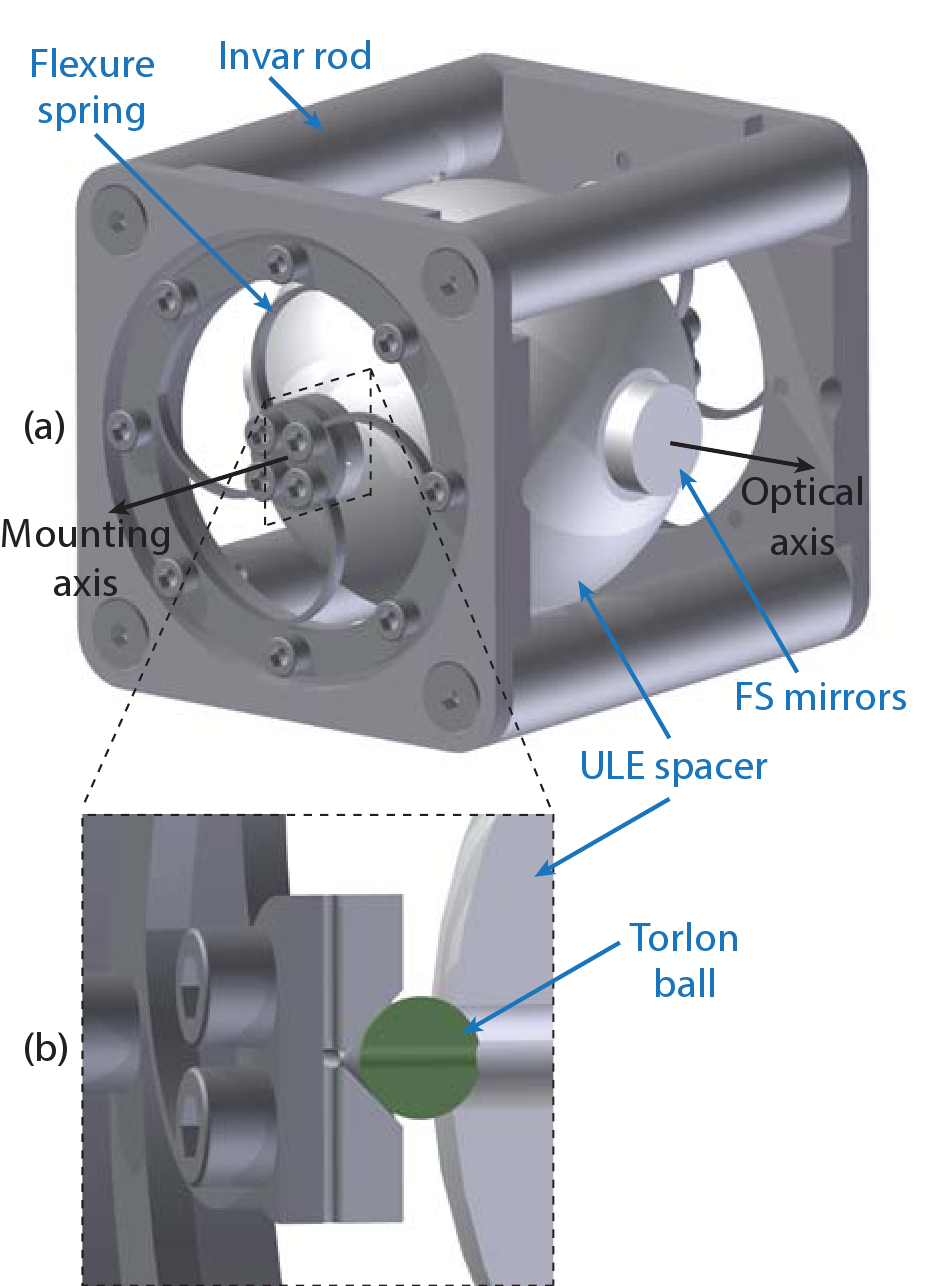}
\caption{\label{fig:cavity}(a) Computer-aided design (CAD) drawing of the mounted Fabry-P\'{e}rot cavity.  The spherical cavity spacer is 50~mm in diameter.  (b) Cross-sectional view showing the details of the contact between the cavity and the mount.  A Torlon ball is compressed between the flexure spring and the vacuum pump-out hole in the cavity.  FS: fused silica.}
\end{center}
\end{figure}

\begin{figure}
\begin{center}
\includegraphics[width=0.9\columnwidth]{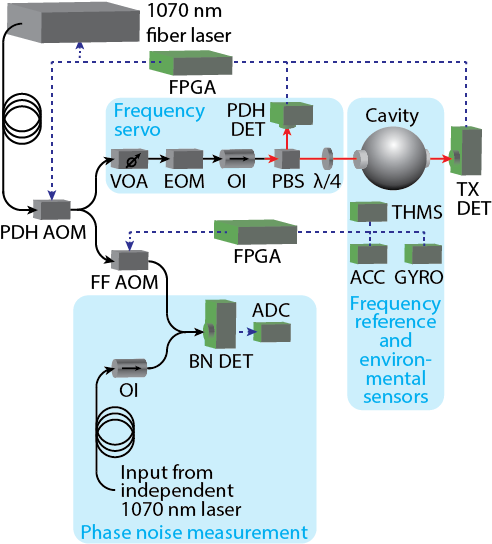}
\caption{\label{fig:schematic}Schematic of the fiber-optic frequency servo and feed-forward.  A laser is stabilized to the length of a spherical Fabry-P\'{e}rot cavity by the PDH method.  Environmental sensors detect inertial forces and temperature drift that perturb the length of the cavity; the resulting frequency perturbations of the laser are corrected by feed-forward to the frequency of an AOM.  The stability of the laser is measured by comparing it with an independent laser.  Note that both the frequency servo and the feed-forward are implemented entirely digitally in FPGAs.  Black lines denote optical fiber, red lines denote free-space laser propagation, and blue dashed lines denote electrical signals.  VOA: variable optical attenuator, OI: optical isolator, PBS: polarizing beam splitter, $\lambda/4$: quarter wave plate, DET: detector, BN: beat-note, TX: transmission, PDH: Pound-Drever-Hall, THMS: thermistor, ACC: accelerometer, GYRO: gyroscope.  See text for other abbreviations.}
\end{center}
\end{figure}

A drawing of the mounted cavity is shown in Fig.~\ref{fig:cavity}a.  The cavity design is similar to that described in Ref.~\cite{Leibrandt2011a}.  Briefly, the spacer is a 50~mm diameter sphere made of Corning Ultra-Low Expansion (ULE) glass \cite{CommercialProduct}.  There is a 6~mm diameter hole drilled along the optical axis and mirrors are optically contacted to 15.2~mm diameter flats polished onto the ends.  A 4~mm diameter hole is drilled orthogonal to the optical axis for vacuum pump-out.  The symmetry of the cavity geometry serves to reduce the acceleration sensitivity \cite{Notcutt2005,Leibrandt2011a}.  The 12.7~mm diameter by 4~mm thick mirrors are made of fused silica for reduced thermomechanical noise \cite{Numata2004,Notcutt2006,Kessler2012a}.  Both mirrors have a radius of curvature of 50~cm and the measured cavity finesse is $7.9(8) \times 10^5$ at the operating wavelength of 1070~nm.

The cavity mount is improved over the one used in Ref.~\cite{Leibrandt2011a}.  Stainless steel 304 flexure springs hold the cavity at two points on a diameter of the spherical spacer that is orthogonal to the optical axis.  The springs are designed such that the rigid-body motional modes of the sphere within the mount are at frequencies of a few hundred Hertz.  This is a compromise between holding the sphere stiffly in order to maintain alignment of the laser to the cavity mode in the presence of $2 g$ acceleration changes, and holding the sphere weakly in order to reduce transmission of stresses due to vibrations and thermal expansion from the mount to the cavity ($1 g$ = 9.8~m/s$^2$).  We calculate the frequencies of the rigid-body motional modes of the cavity using finite-element analysis to be at 280~Hz (translation along the mounting axis), 340~Hz (rotation about the mounting axis), 430~Hz (translation along the orthogonal axes), and 690~Hz (rotation about the orthogonal axes).  For reference, the lowest internal mechanical resonance of the spherical spacer is near 50~kHz.

Contact between the springs and the cavity spacer is made by two 4.8~mm diameter vented Torlon \cite{CommercialProduct} spheres that sit in the hole intended for vacuum evacuation (see Fig.~\ref{fig:cavity}b).  The Torlon spheres are held in place by a $10^2$~N compressive pre-load of the springs.  We find that the residual passive acceleration sensitivity depends primarily on stresses induced in the mounting structure during assembly.  For example, we once disassembled and reassembled the cavity mount without (intentionally) making any changes and the linear acceleration sensitivity changed from $1 \times 10^{-12}$/g, $35 \times 10^{-12}$/g, and $55 \times 10^{-12}$/g to $18 \times 10^{-12}$/g, $7 \times 10^{-12}$/g, and $20 \times 10^{-12}$/g for accelerations along the optical axis, mounting axis, and an orthogonal axis, respectively.  This suggests that in order to make further improvements to the passive acceleration sensitivity, it may be necessary to understand the contact forces between the cavity and the mounting structure in detail.  Improvements may also be gained by use of a mounting structure that can be tuned to null the acceleration sensitivity in-situ \cite{Webster2011}.

The cavity is held at a uniform and stable temperature by housing it in a thermal radiation shield inside a vacuum chamber.  The vacuum chamber and radiation shield are made of aluminum (alloy 7075 for the vacuum chamber and 6061 for the radiation shield) with an electroless nickel coating, and conflat vacuum seals are made using aluminum gaskets \cite{Watanabe1995}.  This design takes advantage of the high thermal conductivity of aluminum and the low emissivity of nickel to ensure a uniform and stable temperature.  The zero crossing of the cavity's coefficient of thermal expansion (CTE) is at approximately -17~$^\circ$C, and we actively stabilize the temperature of the vacuum-chamber wall at 24~$^\circ$C by use of a thermo-electric cooler (TEC).

We lock a 1070~nm fiber laser to the cavity by the Pound-Drever-Hall (PDH) method \cite{Drever1983} (see Fig.~\ref{fig:schematic}).  To improve the long-term stability of the optical alignment, we employ primarily fiber-coupled optics such as a fiber-coupled waveguide electro-optic modulator (EOM) for phase modulation and fiber-coupled acousto-optic modulators (AOMs) \cite{Kessler2012} for closed-loop frequency feedback corrections and open-loop frequency feed-forward corrections.  There is roughly 3~m of optical fiber in the system, and it was not necessary to eliminate phase noise introduced by this optical fiber.  The frequency servo is implemented digitally in a field-programmable gate array (FPGA) that feeds back to the frequency and amplitude of an AOM, a piezo-electric transducer inside the laser, and the temperature of the laser.  Typically, the laser stays locked for days at a time, and the laser optics were adjusted only once during the approximately one-year time period when data in this paper were collected.

We measure inertial forces and temperature drift, which perturb the length of the cavity, both to characterize the environmental sensitivity and to correct for the corresponding frequency perturbations of the laser via feed-forward to the frequency of an AOM.  Linear and rotational accelerations are measured by six one-axis variable-capacitance accelerometers (Dytran 7500A1 \cite{CommercialProduct}) mounted around the cavity in a geometry where there is one sensor at the center of each face of a cube, with the sensing axis oriented on a diagonal of the cube face such that these diagonals form a regular tetrahedron \cite{Chen1994,Leibrandt2011b}.  The accelerometers have a -3~dB bandwidth of 400~Hz.  Rotational velocity is measured by three one-axis MEMS (microelectromechanical systems) gyroscopes (Analog Devices ADIS16130 \cite{CommercialProduct}) with a -3~dB bandwidth of 300~Hz.  Simultaneously with the inertial forces, the temperature of the cavity is measured by a 5~k$\Omega$ thermistor mounted on the outside of the vacuum chamber.  Feed-forward is implemented digitally by use of a second FPGA which controls the frequency of an AOM.

The complete physics package, which includes the laser, cavity, frequency servo and feed-forward optics, and environmental sensors, has a mass of 29~kg and dimensions of $46 \times 46 \times 29$~cm$^3$.  For the broadband acceleration sensitivity measurements described in Sec.~\ref{sec:stability}, the package was mounted on a low-vibration shaker platform.  For the low-frequency acceleration sensitivity measurements described in Sec.~\ref{sec:acceleration}, the physics package was remounted in a structure similar to that of Ref.~\cite{Webster2011}, which allowed the physics package to be rotated around three axes.

\section{Frequency stability and cavity acceleration sensitivity}\label{sec:stability}

\begin{figure}
\begin{center}
\includegraphics[width=0.9\columnwidth]{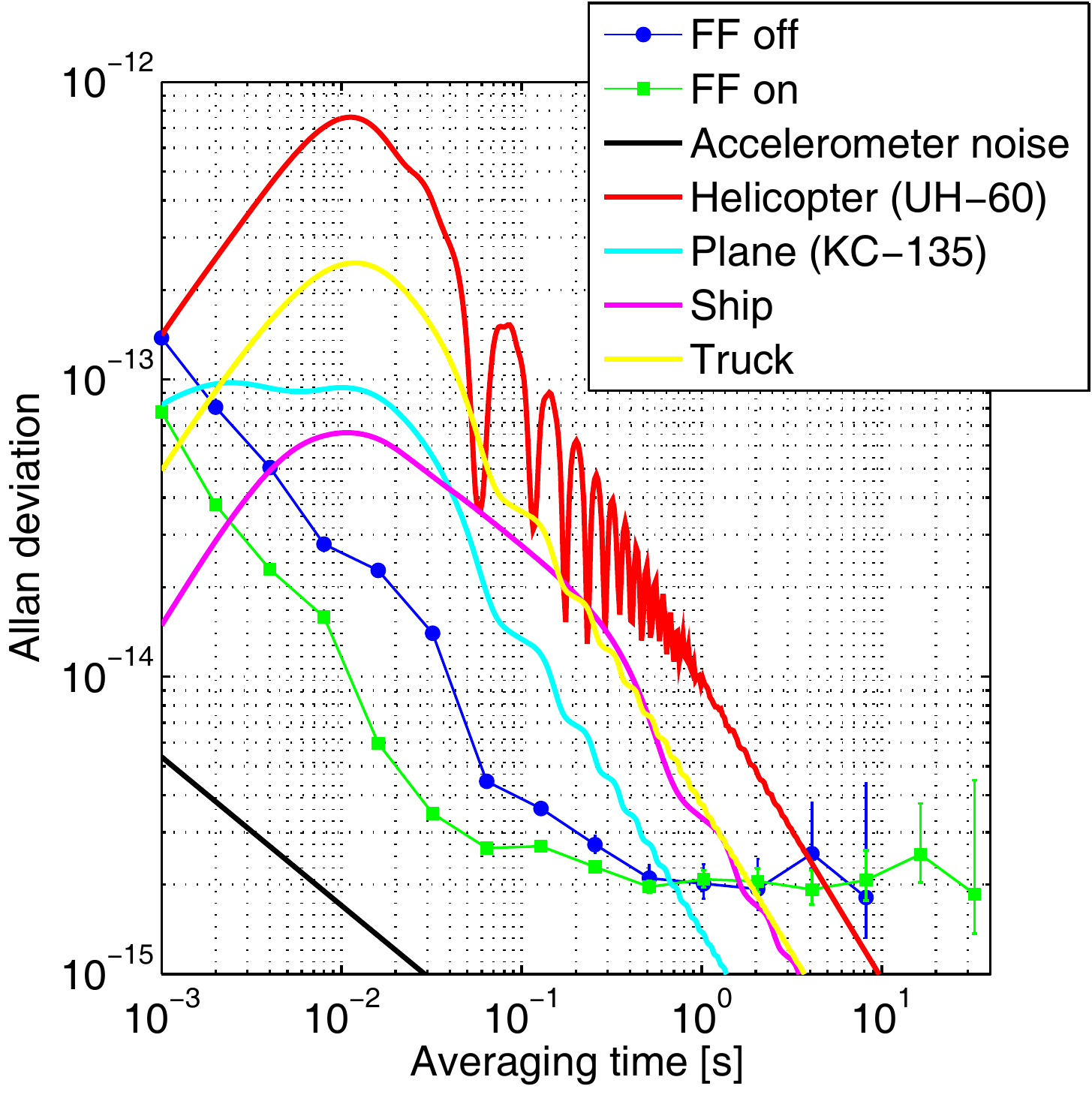}
\caption{\label{fig:AllanDev}Allan deviation of the laser frequency when the physics package is located on a passively isolated optical table, with real-time inertial force feed-forward (FF) off and on.  The data with acceleration feed-forward on is a 120~s data set with no drift removal.  Also shown are the expected contribution of accelerometer noise to laser frequency noise and estimates of the laser instability due to vibrations if a vibration-insensitive laser were to be operated on several types of moving vehicles \cite{MILSTD810G}.  For these estimates, we assume the use of a hypothetical passive vibration isolation system that attenuates accelerations above 30~Hz by 40~dB per decade.  To reach this performance, it will be necessary to extend the feed-forward system, so that the low-frequency acceleration sensitivity of $4 \times 10^{-13}/g$ applies to the entire band of vehicle acceleration frequencies (0--2~kHz).}
\end{center}
\end{figure}

\begin{figure}
\begin{center}
\includegraphics[width=0.9\columnwidth]{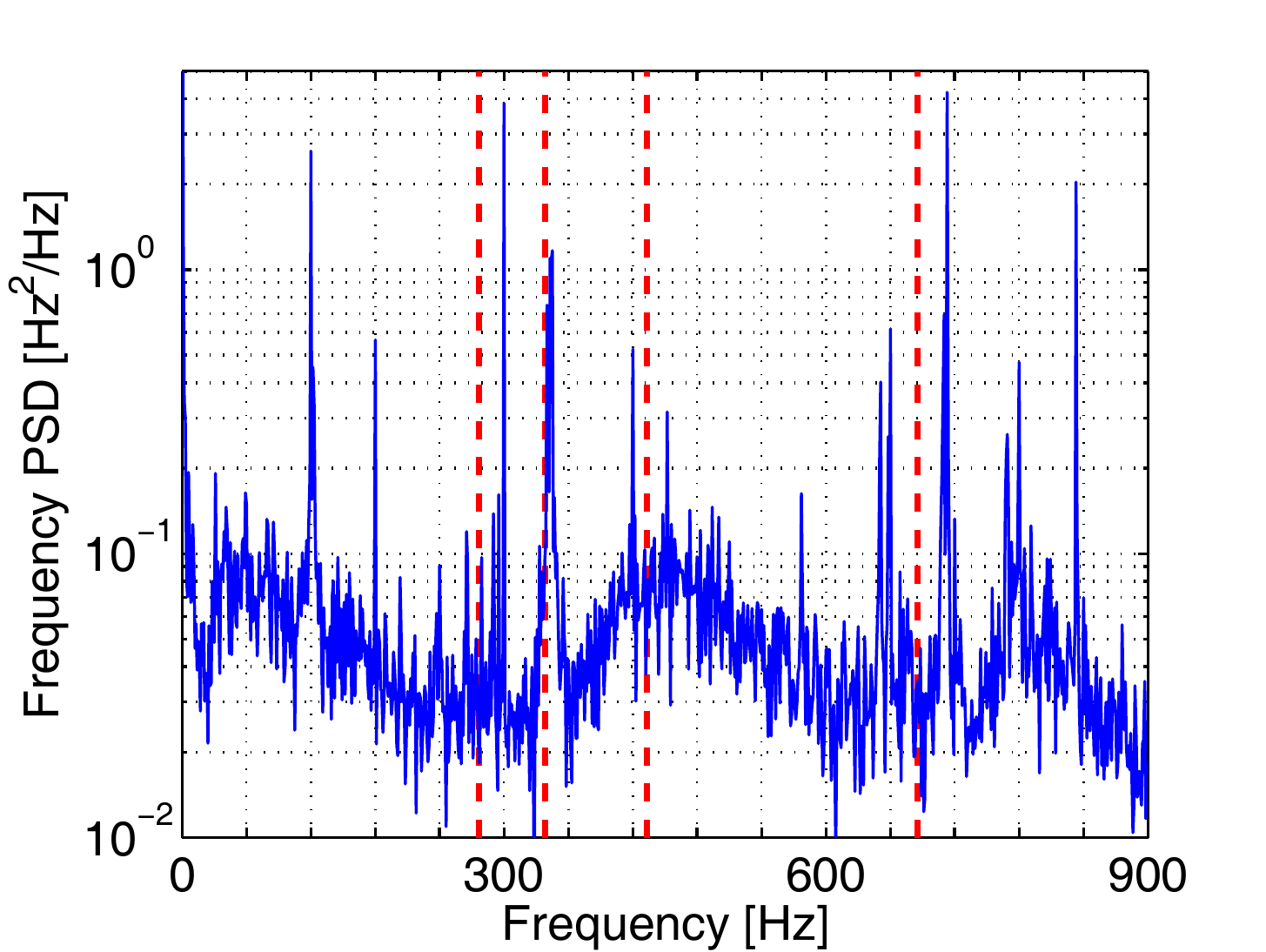}
\caption{\label{fig:freqNoise}Laser frequency power spectral density with the  physics package located on a passively isolated optical table, with real-time inertial force feed-forward on.  The resolution bandwidth is 1~Hz.  Dashed lines denote the calculated frequencies of rigid-body motional resonances of the cavity within its mount.}
\end{center}
\end{figure}

\begin{figure}
\begin{center}
\includegraphics[width=0.9\columnwidth]{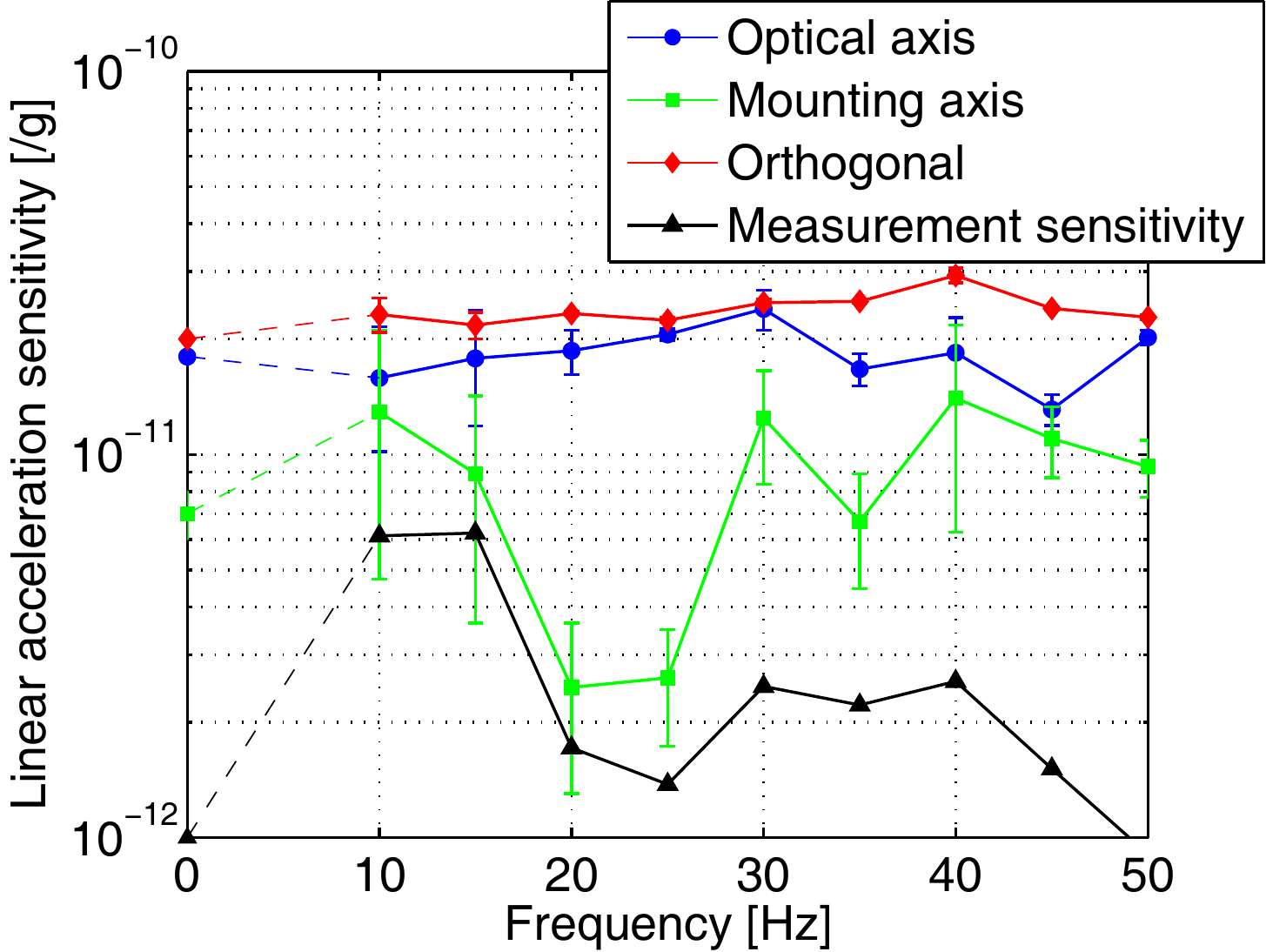}
\caption{\label{fig:ACaccSen}Linear acceleration sensitivity of the cavity versus frequency, without inertial force feed-forward.  These data were taken with the physics package sitting on a commercial active-vibration isolation (AVI) platform operated in reverse as a shaker.  The measurement sensitivity shown in black is limited by the amplitude of vibrations that could be driven by the AVI platform.  With the real-time inertial force feed-forward on, the linear acceleration sensitivity was measured to be consistent with the measurement sensitivity over a frequency range of 0--30~Hz.}
\end{center}
\end{figure}

The frequency stability of the laser is measured via a heterodyne beat note with an independent reference laser that is located in another room.  This reference laser typically serves as the clock laser that drives the $^1$S$_0$ to $^3$P$_0$ transition in $^{27}$Al$^+$ \cite{Chou2010a,Young1999}, and is delivered to the experimental setup via a noise-canceled fiber \cite{Ma1994}.  Figures~\ref{fig:AllanDev} and \ref{fig:freqNoise} show measurements of the laser frequency Allan deviation and power spectral density when the physics package is located on a passively isolated optical table.  The Allan deviation of the laser frequency is $2 \times 10^{-15}$ from 0.5--10~s both with and without applying the inertial force feed-forward corrections (described in Sec.~\ref{sec:acceleration}).  The data show that the fractional frequency fluctuations are reduced by the application of the feed-forward corrections for averaging times shorter than 0.5~s.  Figure~\ref{fig:AllanDev} indicates that accelerometer and gyroscope output noise are not a significant source of laser frequency noise.  The frequency power spectral density shows several narrow spikes at harmonics of the laboratory electric-power frequency (60~Hz) as well as a couple of broader spikes near two of the calculated motional resonance frequencies of the cavity with respect to its mount (340~Hz and 690~Hz).  These spikes do not contain a large fraction of the laser power (both peaks together contain less than 3~\%).

The acceleration sensitivity of the cavity (without inertial force feed-forward) is shown as a function of frequency over 0--50~Hz in Fig.~\ref{fig:ACaccSen}.  This measurement is taken with the cavity on a platform that is actively shaken.  The accelerations and resulting laser frequency noise are simultaneously recorded, and the results are Fourier-transformed and inverted to obtain the acceleration sensitivity for each direction as a function of frequency, as described in Ref.~\cite{Leibrandt2011a}.  The acceleration sensitivity for accelerations along the most sensitive direction (an axis orthogonal to both the optical axis and the mounting axis) is $2(1) \times 10^{-11}$/g over 0--50~Hz.  We do not observe any unexpected mechanical resonances within the measured frequency range.

\section{Inertial force feed-forward}\label{sec:acceleration}

\begin{figure}
\begin{center}
\includegraphics[width=0.9\columnwidth]{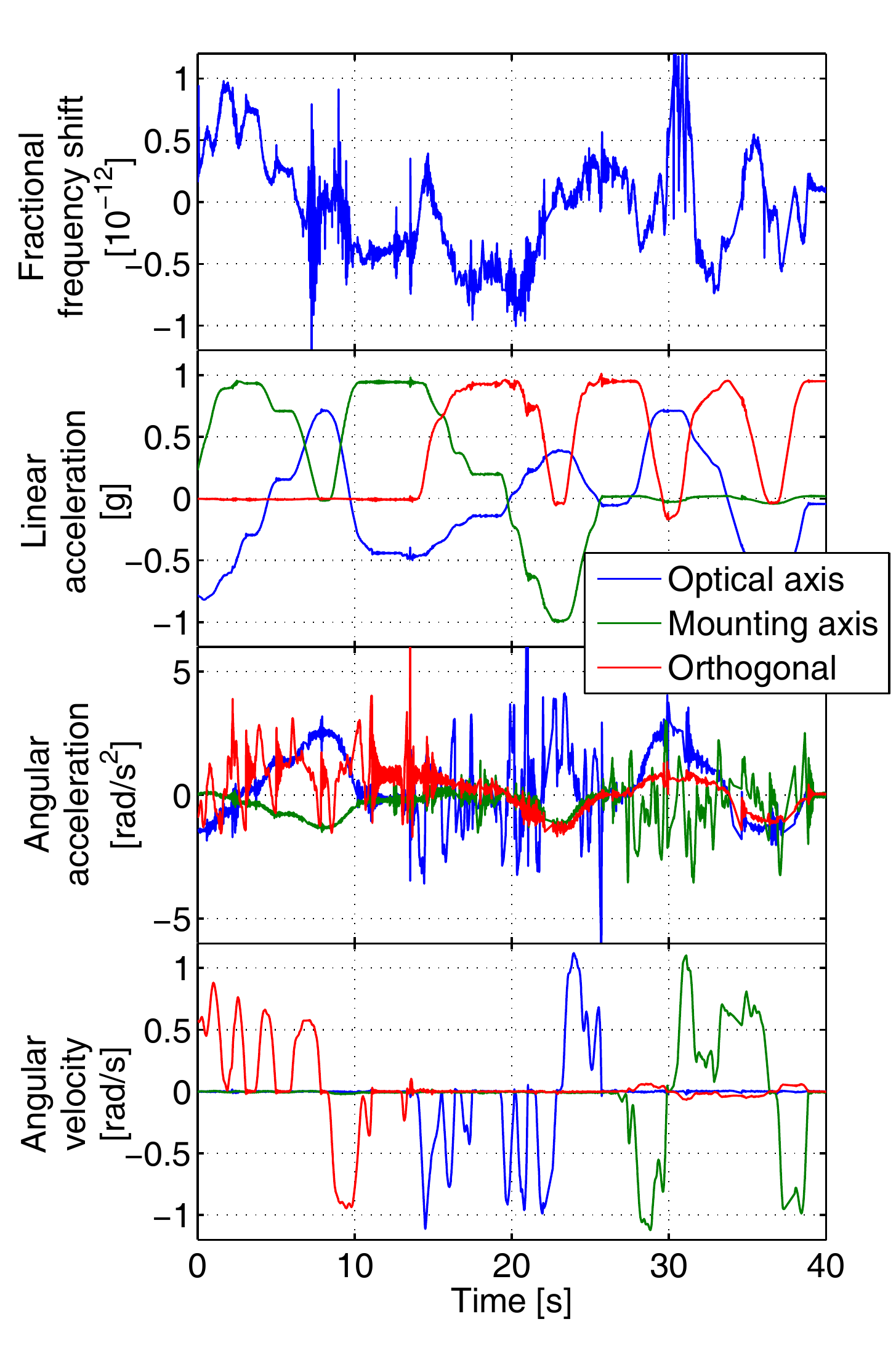}
\caption{\label{fig:DCaccSen}Fractional frequency shift of the laser versus time while rotating the laser and cavity about all three axes, with real-time inertial force feed-forward on.  Also shown from top to bottom are the linear acceleration, angular acceleration, and angular velocity experienced by the laser and cavity.  The data are averaged in 10~ms bins.  The root mean square (RMS) fractional frequency noise of the laser is $4.0 \times 10^{-13}$ and that of the feed-forward corrections is $2.0 \times 10^{-11}$, so inertial force feed-forward has reduced the laser frequency noise power by a factor of 50 in this measurement.  An estimate of the linear acceleration sensitivity with real-time inertial force feed-forward can be obtained by dividing the RMS fractional frequency noise by the RMS acceleration (0.33, 0.47, and 0.73~g along the optical axis, mounting axis, and orthogonal axis) to obtain $4 \times 10^{-13}$/g.  The relative contributions of the different types of inertial forces to the feed-forward frequency correction power are 96.8~\% first order linear acceleration, 2.2~\% second order linear acceleration, 0.1~\% first order rotational acceleration, and 0.9~\% second order rotational velocity.}
\end{center}
\end{figure}

\begin{table*}
\caption{\label{tab:DCaccSen}Sensitivity of the laser frequency to slowly-varying inertial forces.  The origin of the coordinate system in which the inertial forces are measured is coincident with the center of the cavity.  The first group of three lines are without real-time inertial force feed-forward; the second group of three lines are with real-time feed-forward; and the third group of three lines are with real-time feed-forward, measured eight months later using the same Wiener filter taps.  All values should be multiplied by $10^{-12}$ to get the fractional frequency sensitivity.  Units for $\ddot{r}_i$, $\ddot{\theta}_i$, and $\dot{\theta}_i$ are m/s$^2$, rad/s$^2$, and rad/s.  The acceleration due to gravity at the Earth's surface, $g$, is equal to 9.8~m/s$^2$}
\begin{ruledtabular}
\begin{tabular}{rllllll}
				& \multicolumn{2}{c}{Optical axis}					& \multicolumn{2}{c}{Mounting axis}					& \multicolumn{2}{c}{Orthogonal} \\
\hline
Linear acceleration		& $+18(1) (\ddot{r}_x/g)$	& $-1(1) (\ddot{r}_x/g)^2$	& $-7(1) (\ddot{r}_y/g)$		& $+2(1) (\ddot{r}_y/g)^2$	& $+20(1) (\ddot{r}_z/g)$	& $+3(1) (\ddot{r}_z/g)^2$ \\
Rotational acceleration	& $+0.04(2) \ddot{\theta}_x$	&					& $+0.10(2) \ddot{\theta}_x$	&					& $-0.10(2) \ddot{\theta}_x$	& \\
Rotational velocity		&					& $+3.8(2) \dot{\theta}_x^2$	&					& $-6.0(2) \dot{\theta}_y^2$	&					& $-5.3(2) \dot{\theta}_z^2$ \\
\hline
Linear acceleration		& $+0(1) (\ddot{r}_x/g)$		& $+1(1) (\ddot{r}_x/g)^2$	& $+0(1) (\ddot{r}_y/g)$		& $+0(1) (\ddot{r}_y/g)^2$	& $+0(1) (\ddot{r}_z/g)$		& $+0(1) (\ddot{r}_z/g)^2$ \\
Rotational acceleration	& $-0.01(2) \ddot{\theta}_x$	&					& $+0.03(2) \ddot{\theta}_x$	&					& $+0.00(2) \ddot{\theta}_x$	& \\
Rotational velocity		&					& $+0.0(2) \dot{\theta}_x^2$	&					& $+0.3(2) \dot{\theta}_y^2$	&					& $+0.1(2) \dot{\theta}_z^2$ \\
\hline
Linear acceleration		& $+3(1) (\ddot{r}_x/g)$		& $+0(1) (\ddot{r}_x/g)^2$	& $+3(1) (\ddot{r}_y/g)$		& $+2(1) (\ddot{r}_y/g)^2$	& $+0(1) (\ddot{r}_z/g)$		& $+3(1) (\ddot{r}_z/g)^2$ \\
Rotational acceleration	& $+0.02(2) \ddot{\theta}_x$	&					& $+0.00(2) \ddot{\theta}_x$	&					& $+0.02(2) \ddot{\theta}_x$	& \\
Rotational velocity		&					& $-0.2(2) \dot{\theta}_x^2$	&					& $+0.3(2) \dot{\theta}_y^2$	&					& $+0.0(2) \dot{\theta}_z^2$ \\
\end{tabular}
\end{ruledtabular}
\end{table*}

Inertial force feed-forward requires accurate knowledge of the transfer function that describes how linear accelerations, rotational accelerations, and rotational velocities cause the length of the cavity's optical axis to change.  The inertial force sensitivity measurement described in Sec.~\ref{sec:stability} is limited to linear accelerations because our shaker platform cannot drive rotations.  Furthermore, the signal-to-noise ratio of the measurement is low because the maximum amplitude of linear accelerations that can be applied is of the order of $10^{-3}$~g.  In order to improve the accuracy of our knowledge of the transfer function, we remounted the physics package on a platform that allows rotation around three orthogonal axes \cite{Webster2011}.  This enables an accurate measurement of the transfer function for both rotational velocities and accelerations as well as linear accelerations (because the orientation of the cavity changes with respect to gravity), but only at very low frequencies.

First, we perform simultaneous measurements of the accelerometer and gyroscope outputs with the laser frequency while rotating the physics package by hand around three axes.  We use these data to calculate the Wiener filter \cite{Thorpe2010} with one tap for each of the following inputs along each axis: linear acceleration, linear acceleration squared, rotational acceleration, and rotational velocity squared.  The discrete Laplace transform of these taps gives the first- and second-order passive sensitivity of the cavity length to low-frequency linear accelerations, rotational accelerations, and rotational velocity.  The results are listed in the first group of three lines in Tab.~\ref{tab:DCaccSen}.  Note that the second-order sensitivity to linear accelerations is a real effect; the accelerometers we use have a maximum specified nonlinearity of 0.25~\% at 1~g of acceleration, which is smaller than the measured second-order linear acceleration sensitivity.  The measured inertial force transfer function is valid for the frequency range over which we can apply inertial forces by rotating the physics package, which is limited to below 1~Hz.  We expect, however, that the inertial force transfer function is constant in frequency up to at least 50~Hz because we do not see any mechanical resonances in that frequency range (see Fig.~\ref{fig:ACaccSen}).

After calibrating the Wiener filter taps we again rotate the physics package, while performing real-time inertial force feed-forward.  Simultaneously, we acquire a second set of measurements of the accelerometer and gyroscope outputs together with the laser frequency.  From these data we obtain the inertial force sensitivity of the laser frequency after real-time feed-forward correction.  The results, which are shown in the second group of three lines in Tab.~\ref{tab:DCaccSen}, are consistent with zero within our measurement sensitivity, which is $10^{-12}$/g for linear accelerations.  In other words, we are able to actively correct the inertial force sensitivity as well as we can measure it.

Figure~\ref{fig:DCaccSen} shows an example measurement of the inertial force sensitivity with real-time inertial force feed-forward running, obtained by rotating the laser about three axes.  In this dataset, the laser frequency fluctuations due to inertial forces are reduced by a factor of 50 to a root mean square (RMS) fractional amplitude of $4.0 \times 10^{-13}$.  We can get an upper bound of the first-order linear acceleration sensitivity from this data, assuming that the acceleration sensitivity is isotropic and that the accelerations applied along the three axes are linearly independent, by dividing the RMS fractional frequency noise by the RMS acceleration (0.33, 0.47, and 0.73~g along the optical axis, mounting axis, and orthogonal axis) to obtain $4 \times 10^{-13}$/g.  This is consistent with the values reported in the second group of three lines in Tab.~\ref{tab:DCaccSen}.  Figure~\ref{fig:AllanDev} includes an estimate of the laser frequency noise due to accelerations if a laser with an acceleration sensitivity of $4 \times 10^{-13}$/g were to be operated in several varieties of moving vehicles.

In order to investigate the long-term stability of the acceleration sensitivity, we remeasured the acceleration sensitivity 8 months after the first measurement with real-time inertial force feed-forward running using the same Wiener filter taps as the first measurement.  The results are listed in the third group of three lines in Tab.~\ref{tab:DCaccSen}.  The linear acceleration sensitivity changed by $3 \times 10^{-12}$/g for accelerations along two directions.  We have verified that this is due to a change of the passive acceleration sensitivity of the cavity, which may be caused by material creep of the cavity or mounting structure.  Long-term operation of this system with an acceleration sensitivity below $10^{-12}$/g would thus require periodic recalibration of the Wiener filter taps.

\section{Temperature feed-forward}\label{sec:temperature}

\begin{figure}
\begin{center}
\includegraphics[width=0.9\columnwidth]{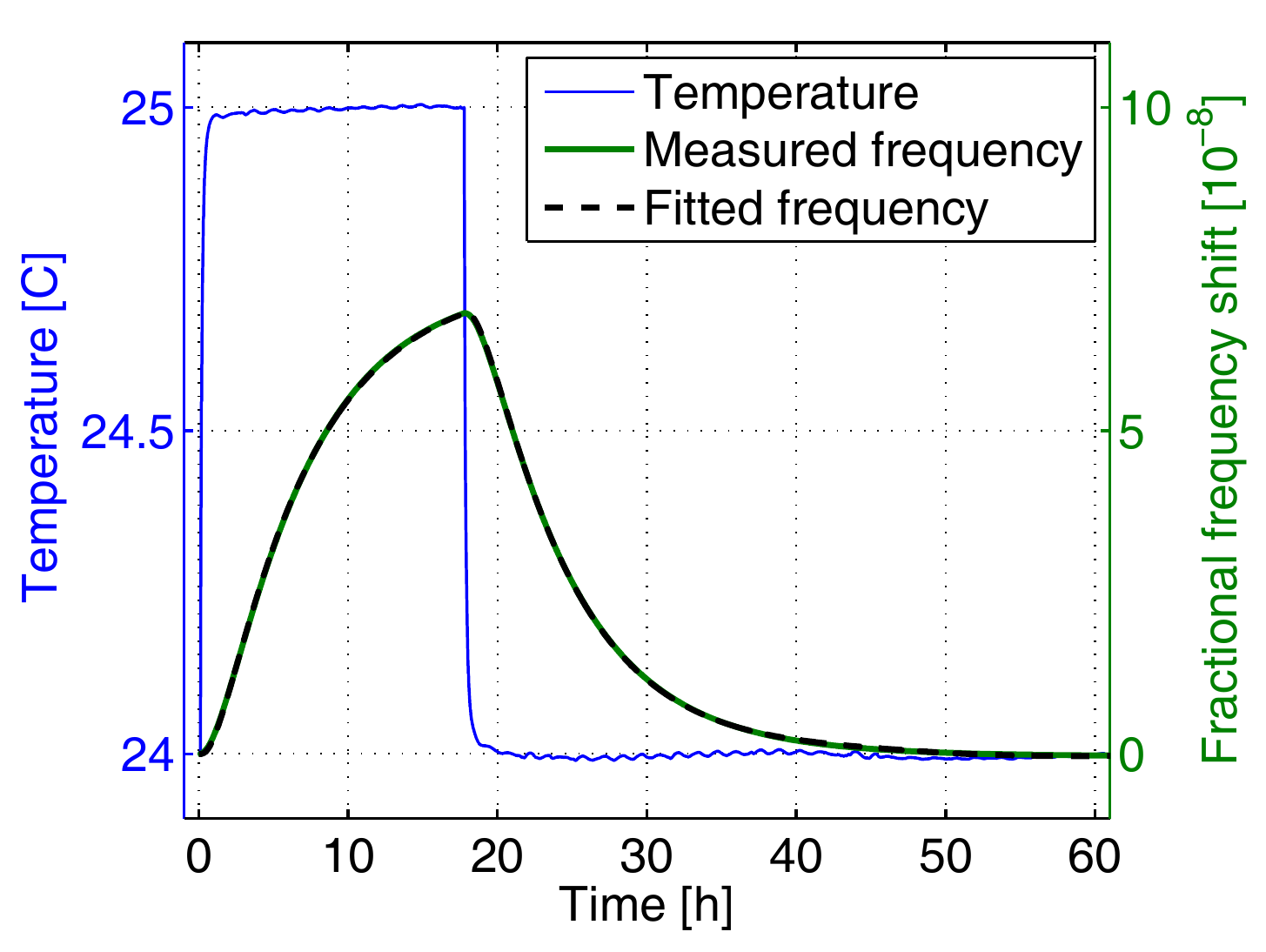}
\caption{\label{fig:tempStep}Fractional frequency shift of the laser (green line) while the temperature of the vacuum chamber which holds the cavity (blue line) is stepped by 1~K.  The fit (black dashed line) is a second order low-pass filter of the measured temperature with time constants of 1.1 and 6.1~h, scaled by $7.3 \times 10^{-8}$/K.}
\end{center}
\end{figure}

We measure the transfer function from temperature fluctuations of the environment to length changes of the cavity by simultaneously recording the temperature of the vacuum chamber and the frequency of the laser while making a step in the setpoint of the vacuum-chamber temperature controller.  The results are shown in Fig.~\ref{fig:tempStep}.  The measured laser frequency is fitted to a second-order low-pass filter of the measured temperature with time constants of 1.1 and 6.1~h, scaled by $7.3 \times 10^{-8}$/K.  We use this temperature sensitivity in a finite element model of the cavity to determine that the zero crossing of the cavity's CTE is at approximately -17~$^\circ$C \cite{Fox2009,Legero2010}.  The time constants are in reasonable agreement with estimates of the thermal conductivity between the vacuum chamber and the radiation shield and between the radiation shield and the cavity.

\begin{figure}
\begin{center}
\includegraphics[width=0.9\columnwidth]{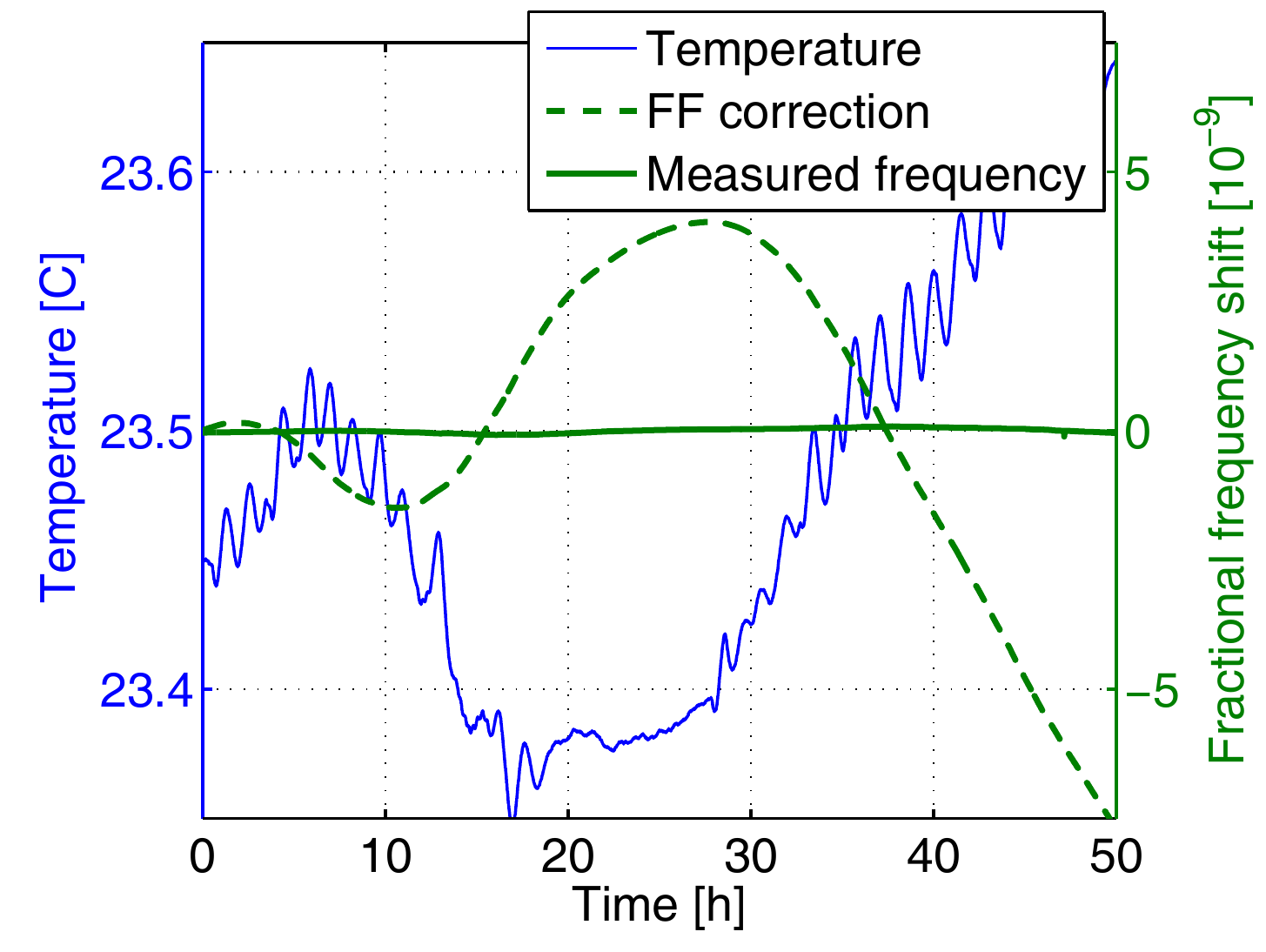}
\caption{\label{fig:tempFF}Temperature of the vacuum chamber that holds the cavity (blue line) and the corresponding fractional frequency shift of the laser (green line) while the temperature of the vacuum chamber is measured and thermal length changes of the cavity are compensated by feed-forward to the frequency of an acousto-optic modulator (AOM).  Temperature control of the vacuum chamber is disabled during this measurement, so the vacuum chamber temperature drifts with the temperature of the laboratory.  The root mean square (RMS) fractional frequency shift of the laser is $4.5 \times 10^{-11}$ and that of the feed-forward corrections is $3.0 \times 10^{-9}$, so temperature feed-forward has reduced the laser frequency noise by a factor of 70 in this measurement.}
\end{center}
\end{figure}

\begin{figure}
\begin{center}
\includegraphics[width=0.9\columnwidth]{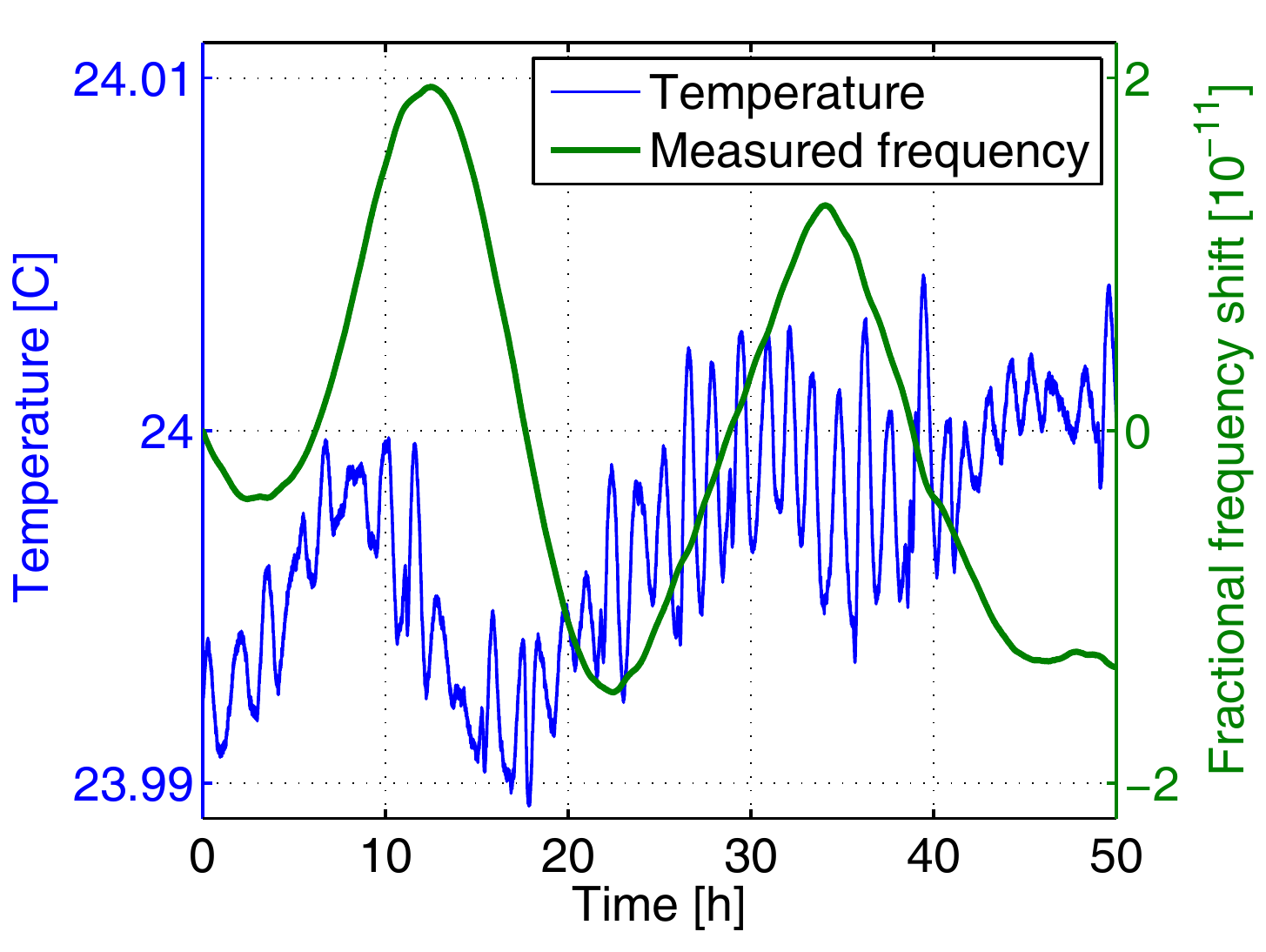}
\caption{\label{fig:tempFB}Temperature of the vacuum chamber that holds the cavity (blue line) and the corresponding fractional frequency shift of the laser (green line) while the temperature of the vacuum chamber is stabilized.  The root mean square (RMS) fractional frequency shift of the laser is $1.0 \times 10^{-11}$.}
\end{center}
\end{figure}

We implement real-time temperature feed-forward using the transfer function measured above, discretized with a 1.1~s time step.  Note that we position the temperature sensor on the outside of the vacuum chamber intentionally, rather than on the radiation shield or on the cavity itself, because in this configuration the sensor's noise is heavily low-pass filtered by the long radiative time constants before it is added onto the laser frequency.  Figure~\ref{fig:tempFF} shows a measurement in which the temperature controller is disabled, and the cavity temperature drifts with the laboratory temperature.  Real-time temperature feed-forward reduces the laser frequency shift by a factor of 70, to an RMS fractional frequency shift of $4.5 \times 10^{-11}$ in an environment where the temperature of the vacuum chamber drifts by 80~mK RMS.  These residual frequency fluctuations can be compared with the RMS factional frequency shift with the temperature controller enabled shown in Fig.~\ref{fig:tempFB}, which is $1.0 \times 10^{-11}$.  While the performance of temperature feed-forward is worse than temperature control (i.e., feedback) in this implementation, this trade-off might be acceptable for some applications where power is limited (e.g., field \cite{Leibrandt2011a} or cryogenic \cite{Thorpe2011,Kessler2012} systems).  Temperature feed-forward might also be improved by using multiple or more precise temperature sensors.

\section{Conclusion}

We have demonstrated, by a combination of an improved cavity-mount design and real-time feed-forward that corrects for all of the inertial forces to second-order, a cavity-stabilized laser with acceleration sensitivity below $10^{-12}$/g.  The performance of inertial force feed-forward in this work is limited, at least in part, by our inability to measure the transfer function with sufficient accuracy.  It may be possible to obtain more complete knowledge of the transfer function, and thus better feed-forward reduction of the acceleration sensitivity, with the use of a shaker table capable of applying broadband, six degree-of-freedom, large-amplitude accelerations.  In addition, we have shown that feed-forward can also be used to reduce laser frequency noise due to temperature fluctuations by use of a temperature sensor located on the outside of the vacuum chamber.  These two examples illustrate the power and generality of feed-forward for laser frequency stabilization: laser frequency noise caused by environmental fluctuations can be strongly reduced, as long as the environmental variable can be precisely measured and the transfer function accurately known.

\section*{ACKNOWLEDGMENTS}

We thank R.~Drullinger, M.~Notcutt, and M.~Thorpe for their contributions to earlier work on low acceleration sensitivity lasers and also for useful discussions; R.~Lalezari for fabrication of the cavity used in this work; and A.~Ludlow and W.~Swann for critical readings of this manuscript.  This work is supported by AFOSR, DARPA, and ONR and is not subject to US copyright.


\end{document}